\documentclass[iop]{emulateapj}

\usepackage{epsfig}
\usepackage{natbib}
\usepackage{graphicx}
\usepackage{multirow}
\usepackage{float}
\usepackage{hyperref}
\usepackage{amsmath}
\def\simgt{\lower.5ex\hbox{$\; \buildrel > \over \sim \;$}}
\def\simlt{\lower.5ex\hbox{$\; \buildrel < \over \sim \;$}}

\def\amin{\ifmmode^{\prime}\else$^{\prime}$\fi}
\def\asec{\ifmmode^{\prime\prime}\else$^{\prime\prime}$\fi}

\def\simgt{\lower.5ex\hbox{$\; \buildrel > \over \sim \;$}}
\def\simlt{\lower.5ex\hbox{$\; \buildrel < \over \sim \;$}}

\newcommand\chandra{{\it Chandra}}

\newcommand\xmm{{\it XMM-Newton}}

\newcommand\swift{{\it Swift\/}}
\newcommand\nustar{{\it NuSTAR\/}}

\newcommand\spitzer{{\it Spitzer\/}}

\def\sgra{Sgr~A$^{\star}$}

\shorttitle{\sgra\ X-ray Flares}
\shortauthors{S. Zhang et~al.}

\begin{document}

\title{Sagittarius A$^{\star}$ High Energy X-ray Flare Properties \\ During \nustar\ Monitoring of the Galactic Center From 2012 to 2015}
        
\author{Shuo Zhang\altaffilmark{1, 2}, Frederick K. Baganoff\altaffilmark{1}, Gabriele Ponti\altaffilmark{3}, Joseph Neilsen\altaffilmark{1}, 
John A. Tomsick\altaffilmark{4}, \\ Jason Dexter\altaffilmark{3}, Ma{\"i}ca Clavel\altaffilmark{4}, Sera Markoff\altaffilmark{5}, Charles J. Hailey\altaffilmark{2}, Kaya Mori\altaffilmark{2}, \\ Nicolas M. Barri{\`e}re\altaffilmark{4}, 
Michael A. Nowak\altaffilmark{1}, Steven E. Boggs\altaffilmark{4}, Finn E. Christensen\altaffilmark{6}, William W. Craig\altaffilmark{4,7}, \\ Brian W. Grefenstette\altaffilmark{8}, Fiona A. Harrison\altaffilmark{8}, Kristin K. Madsen\altaffilmark{8}, Daniel Stern\altaffilmark{9}, William W. Zhang\altaffilmark{10}}

\altaffiltext{1}{MIT Kavli Institute for Astrophysics and Space Research, Cambridge, MA 02139, USA; shuo@mit.edu}
\altaffiltext{2}{Columbia Astrophysics Laboratory, Columbia University, New York, NY 10027, USA}
\altaffiltext{3}{Max-Planck-Institut f\"{u}r extraterrestrische Physik, Giessenbachstrasse 1, D-85748, Garching bei M\"{u}nchen, Germany}
\altaffiltext{4}{Space Sciences Laboratory, University of California, Berkeley, CA 94720, USA}
\altaffiltext{5}{Astronomical Institute,``Anton Pannekoek", University of Amsterdam, Postbus 94249, 1090 GE Amsterdam, The Netherlands}
\altaffiltext{6}{DTU Space - National Space Institute, Technical University of Denmark, Elektrovej 327, 2800 Lyngby, Denmark}
\altaffiltext{7}{Lawrence Livermore National Laboratory, Livermore, CA 94550, USA}
\altaffiltext{8}{Cahill Center for Astronomy and Astrophysics, California Institute of Technology, Pasadena, CA 91125, USA}
\altaffiltext{9}{Jet Propulsion Laboratory, California Institute of Technology, Pasadena, CA 91109, USA}
\altaffiltext{10}{X-ray Astrophysics Laboratory, NASA Goddard Space Flight Center, Greenbelt, MD 20771, USA}

\begin{abstract}

Understanding the origin of the flaring activity from the Galactic center supermassive black hole, Sagittarius~A$^{\star}$, is a major scientific goal of the \nustar\ Galactic plane survey campaign.
We report on the data obtained between July 2012 and April 2015, including 27 observations on \sgra\ with a total exposure of $\simeq1$ Ms.
We found a total of ten X-ray flares detected in the \nustar\ observation window, with luminosities in the range of $L_{3-79\rm~keV}\sim(0.2$--$4.0)\times10^{35}\rm~erg~s^{-1}$. 
With this largest hard X-ray \sgra\ flare dataset to date, we studied the flare spectral properties.
Seven flares are detected above $5\sigma$ significance,    
showing a range of photon indices ($\Gamma \sim 2.0$--2.8) with typical uncertainties of $\pm 0.5$ (90\% confidence level).
We found no significant spectral hardening for brighter flares as indicated by a smaller sample.
The accumulation of all the flare spectra in 1--79~keV can be well fit with an absorbed power-law model with $\Gamma=2.2\pm0.1$, and does not require the existence of a spectral break.   
The lack of variation in X-ray spectral index with luminosity would point to a single mechanism for the flares and is consistent with the synchrotron scenario.
Lastly, we present the quiescent state spectrum of \sgra, and derived an upper limit on the quiescent luminosity of \sgra\ above 10~keV to be $L_{Xq, 10-79\rm~keV} \le (2.9\pm0.2)\times10^{34}$~erg~$\rm s^{-1}$. 

\end{abstract}


\section{Introduction}

Sagittarius~A$^{\star}$ (\sgra), located at the Galactic nucleus of the Milky Way Galaxy, is one of the most underluminous supermassive black holes (SMBH) known.
The current quiescent bolometric luminosity of \sgra\ is $L\simeq10^{36} \rm erg~s^{-1}$, which is roughly eight orders of magnitude lower than its Eddington luminosity of a $4 \times 10^{6} M_\sun$ black hole \citep{Narayan1998, Ghez2008}.
However, there has been observational evidence indicating that \sgra\ could have been much brighter in the past (e.g. \citealp{Zhang2015, Ponti2013} and the reference therein).  
As the closest SMBH to Earth \citep{Reid2004}, \sgra\ is an ideal laboratory to study accretion processes of quiescent black hole systems \citep{FalckeMarkoff2013}.

The X-ray emission of its quiescent state comes from an optically thin thermal plasma with $kT\sim 2$~keV that extends out to the Bondi radius about $10^{5}$ times the gravitational radii ($r_{B} \sim 10^{5} r_{g}$) \citep{Quataert2002, Baganoff2003, Wang2013}.
The X-ray quiescent state of \sgra\ is punctuated by flares lasting up to a few hours (e.g. \citealp{Baganoff2001, Porquet2003, Dodds-Eden2009, Trap2011, Neilsen2013, Neilsen2015, Degenaar2013, Barriere2014, Ponti2015}).   
During the flares, the X-ray luminosity of \sgra\ increases by a factor of up to a few hundred over the quiescent level \citep{Porquet2003, Nowak2012}.
Fast variability with timescales of a few hundred seconds \citep{Porquet2003, Nowak2012, Barriere2014} suggests a compact emission region within a few gravitational radii from the black hole ($r_{g}/c=20\rm~s$).
Therefore, flares hold the key to probe the physical conditions in the immediate vicinity of the SMBH.

After a decade of intense \sgra\ monitoring, there still remain many puzzles regarding the origin of the flaring activity (e.g.\ see review by \citealp{Genzel2010}). 
Two distinctively different classes of models have been proposed as the origin of the flares: electron acceleration processes \citep{Markoff2001, Liu2004,Yuan2004, Dodds-Eden2010, Dibi2014}, and transient events in the \sgra\ accretion flow \citep{TaggerMelia2006, BroderickLoeb2005, Yusef-Zadeh2006, Eckart2006a, Trap2011, Zubovas2012}.
The flare models mentioned above invoke two types of radiation meachanisms for the X-ray flares: 1) synchrotron emission (with cooling break, or SB model) where the NIR to the X-ray emission is generated from one population of electrons; 2) inverse Compton (IC) emission where the NIR emitting electrons up-scatter the NIR synchrotron emission itself (i.e. synchrotron self-Compton, SSC) or the sub-mm photons from the environment (external Compton, EC).
Recent multi-wavelength observations of a bright \sgra\ flare indicate synchrotron emission with a cooling break and an evolving high energy cut-off as the most likely mechanism \citep{Ponti2017}.

Dozens of \sgra\ X-ray flares have been observed so far, mainly by \chandra, \xmm\ and \swift.
As different flare radiation models predict different spectral shapes, the spectral properties of these flares carry vital information for us to understand the radiation mechanisms and ultimately the physical processes behind the flares.
Recent studies discussed whether the flare spectral shapes depend on the luminosities \citep{Porquet2003, Nowak2012, Degenaar2013}.
During the \chandra\ \sgra\ X-ray Visionary Project (XVP), thirty-nine X-ray flares were detected in 2--8~keV \citep{Neilsen2013}. 
Data in this relatively narrow bandwidth did not provide evidence for X-ray color differences between faint and bright flares.
The analysis of the \xmm\ data confirms this result in the 3--10~keV energy band; however, it suggests spectral evolution within each flare \citep{Ponti2017}. 
 
The flare spectrum beyond 10~keV has the potential to help distinguish between the synchrotron-type model (which predicts a single power-law spectrum) and the IC-type model (which instead predicts an X-ray spectrum with curvature).
Using the 3--79~keV data obtained by \nustar\ in 2012, \citet{Barriere2014} for the first time reported different spectral indices between two flares, with a harder spectrum detected for the brighter flare at 95\% confidence level.
However, due to limited statistics and a limited number of flares, neither emission mechanism could be ruled out. 
While the SB model has been preferred for its more physical parameters \citep{Dodds-Eden2009, Barriere2014}, \citet{Dibi2016} shows some challenges to this model through the first statistical study of flare models using \chandra\ observations. 
More X-ray flares detected in the broad X-ray band with good statistics need to be accumulated to answer these unsolved questions.

Aiming at building a large database of X-ray flares of different luminosities, durations, and spectra, \nustar\ has been monitoring \sgra\ through the Galactic Center observing campaign since its launch in 2012. 
In this paper we report on the \nustar\ Galactic Center observing campaign, and our \sgra\ flare study results using data obtained from 2012 to 2015.
We searched for X-ray flares from all 27 Galactic Center observations with \sgra\ in the field of view (FoV), totaling $\sim1$~Ms of exposure time.
Besides the four flares reported in \citet{Barriere2014}, six more \sgra\ hard X-ray flares were detected, resulting a total of ten \nustar\ flares, seven simultaneously detected by \chandra\ or \xmm.
Using the largest broadband X-ray flare database by far, we investigated the spectral properties for all the flares.
The paper is organized as follows.
In Section 2, we introduce the \nustar\ Galactic Center observation campaign.
In Section 3, we present the data reduction.
We demonstrate the flare search results in Section 4. 
In section 5, we present the spectral properties for \sgra\ flares and quiescent state, which are discussed in Section 6.

\section {\nustar\ Galactic Center Observing Campaign}

\sgra\ is a key target of the \nustar\ Galactic Center campaign.  
The first \sgra\ observation was initiated in 2012 July as a coordinated observation campaign with \chandra\ and Keck.  
Three \nustar\ Galactic Center observations resulted in 375~ks total exposure time, during which four bright flares with X-ray luminosity in the range of  $L_{3-79\rm~keV}=(0.73-3.97)\times 10^{35}\rm erg~s^{-1}$ were detected by \nustar\ up to 79~keV \citep{Barriere2014}. 
The bright flare detected in October 2012 was simultaneously detected by \chandra, while no X-ray flare was covered by the Keck observation window.
The \sgra\ region was also covered by four out of six pointings ($\sim25$~ks exposure each) of the \nustar\ Galactic Center mini-survey conducted in October 2012 \citep{Mori2015}.

In 2013, major X-ray observatories, including \chandra, \xmm\ and \swift, conducted long \sgra\ observing campaigns in order to investigate potential variation in \sgra\ X-ray activity caused by pericenter passage of the very red Br$\gamma$ object called G2 \citep{Gillessen2012, Witzel2014}. 
A recent study of all 150 \xmm\ and \chandra\ Galactic Center observations over the last 15 years reports a significant increase in the number and average luminosity of bright flares happening after the pericenter passage of G2 \citep{Ponti2015}.
It is still uncertain whether this variation is due to clustering of bright flares observed during more frequent monitoring or increased accretion activity induced by G2.
The outburst of SGR~J1745$-$29 \citep{Kennea2013, Mori2013, Rea2013}, a transient magnetar only 2.4\asec\ from \sgra, triggered further observations of the Galactic Center region in 2013.   
Later in 2013, two X-ray transients, CXOGC~J174540.0$-$290005 and AXJ~1745.6$-$2901, went into outburst at different times (see ATELs 5095, 5074, 5226, 1513).
\nustar\ allocated a total of $\sim380$~ks to monitor these Galactic Center transient phenomena in 2013.
These observations were dominated by the bright X-ray transients, thus making it impossible for \nustar\ to characterize even the brightest \sgra\ flares.

As the magnetar SGR~J1745$-$29 became less dominant, another 100~ks \nustar\ observation was allocated to a multi-wavelength \sgra\ observation campaign coordinated with \chandra\ and \spitzer\ in summer 2014. 
A third multi-wavelength campaign (\nustar\, \xmm\, SINFONI-VLT and VLBA) was performed after the pericenter passage of G2 (see \citealp{Ponti2017}).
A summary of all the 27 \nustar\ observations with \sgra\ in the FoV are listed in Table \ref{tab:SgrAobs}.

\begin{deluxetable*}{lccccccc}
\tablecaption{\nustar\ Galactic Center Observations During 2012 to 2015 and Simultaneous X-ray Observations} 
\tablecolumns{7}                                                                                                                                                                                              
\tablehead{\colhead{\nustar\ Obs} &\colhead{ } &\colhead{ } &\colhead{ } &\colhead{Joint Obs} &\colhead{ } &\colhead{ } &\colhead{ } \\ 
\colhead{Target} &\colhead{obsID} &\colhead{Start(UTC)} &\colhead{Exp} &\colhead{Instrument} & \colhead{obsID} & \colhead{Start(UTC)}  &\colhead{Exp} }
\startdata
\sgra\ 					& 30001002001	  & 2012-07-20 02:11:07 	  &154.2~ks	  & \chandra\  		&13842   		& 2012-07-21 11:52:48 		& 191.7~ks \\
\sgra\ 					& 30001002003  	  & 2012-08-04 07:56:07  	  &77.1~ks       & \chandra\  		&13852   		& 2012-08-04 02:36:57 		& 156.6~ks \\
\sgra\ 					& 30001002004  	  & 2012-10-16 18.31:07   	  &49.6~ks    	  & \chandra\  		& 13851   		& 2012-10-16 18:48:57 		& 107.1~ks \\
Mini-survey 				& 40010001002  	  & 2012-10-13 06:41:07  	  &23.9~ks    	  & \nodata     		& \nodata 		& \nodata                    		& \nodata      \\
Mini-survey 				& 40010002001  	  & 2012-10-13 19:21:07  	  &24.2~ks    	  & \nodata     		& \nodata 		& \nodata                    		& \nodata      \\
Mini-survey 				& 40010003001  	  & 2012-10-14 09:56:07  	  &24.0~ks    	  & \nodata     		& \nodata 		& \nodata                    		& \nodata      \\
Mini-survey 				& 40010004001  	  & 2012-10-15 00:31:07  	  &24.0~ks    	  & \nodata     		& \nodata 		& \nodata                    		& \nodata      \\
SGR~J1745$-$29 			& 30001002006 	  & 2013-04-26 01:01:07 	  &37.2~ks 	  & \nodata 		& \nodata 		& \nodata 					& \nodata 	  \\
SGR~J1745$-$29 			& 80002013002 	  & 2013-04-27 06:16:07	  &49.8~ks 	  & \nodata 		& \nodata 		& \nodata 					& \nodata 	  \\
SGR~J1745$-$29 			& 80002013004 	  & 2013-05-04 17:31:07  	  &38.6~ks 	  & \nodata 		& \nodata 		& \nodata 					& \nodata 	  \\
SGR~J1745$-$29 			& 80002013006 	  & 2013-05-11 14:26:07 	  &32.7~ks  	  & \nodata 		& \nodata 		& \nodata 					& \nodata 	  \\
SGR~J1745$-$29 w/T1* 		& 80002013008 	  & 2013-05-18 17:36:07  	  &39.0~ks  	  & \nodata 		& \nodata 		& \nodata 					& \nodata 	  \\
SGR~J1745$-$29 w/T1 		& 80002013010 	  & 2013-05-27 10:16:07   	  &37.4~ks  	  & \nodata 		& \nodata 		& \nodata 					& \nodata 	  \\
SGR~J1745$-$29 			& 80002013012 	  & 2013-06-14 09:56:07	  &26.7~ks  	  & \nodata 		& \nodata 		& \nodata 					& \nodata 	  \\
SGR~J1745$-$29 			& 80002013014/6  	  & 2013-06-07 04:16:07 	  &29.5~ks 	  & \nodata 		& \nodata 		& \nodata 					& \nodata 	  \\
SGR~J1745$-$29 w/T2** 		& 80002013018 	  & 2013-07-31 01:56:07	  &22.3~ks 	  & \nodata 		& \nodata 		& \nodata 					& \nodata 	  \\
SGR~J1745$-$29 w/T2 		& 80002013020 	  & 2013-08-08 15:01:07 	  &12.0~ks 	  & \nodata 		& \nodata 		& \nodata 					& \nodata 	  \\
SGR~J1745$-$29 w/T2 		& 80002013022 	  & 2013-08-09 09:01:07 	  &11.2~ks 	  & \nodata 		& \nodata 		& \nodata 					& \nodata 	  \\
SGR~J1745$-$29 w/T2 		& 80002013024 	  & 2013-08-13 00:06:07	  &11.7~ks 	  & \nodata 		& \nodata 		&\nodata 					& \nodata 	  \\
\sgra\ w/T2 				& 30001002008  	  & 2014-06-18 02:21:07     &33.1~ks    	  & \nodata   		& \nodata          & \nodata          	                 & \nodata      \\
\sgra\ w/T2 				& 30001002010  	  & 2014-07-04 10:36:07  	  &61.3~ks    	  & \chandra  		& 16597 		& 2014-07-05 02:14:47   	        & 16.5~ks      \\
\sgra\ w/T2			     	& 30002002002     	  & 2014-08-30 19:45:07  	  &59.8~ks    	  & \xmm\       		& 0743630201 & 2014-08-30 19:20:01 		& 33.9~ks 	  \\
         				        & 				  & 					  &            		  & \xmm\       		& 0743630301 & 2014-08-31 20:23:30 		& 26.9~ks 	  \\
          					& 				  & 					  &          		  & \chandra\  		& 16217           & 2014-08-30 04:49:05 		& 34.5~ks 	  \\                
\sgra\ w/T2				& 30002002004  	  & 2014-09-27 17:31:07     &67.2~ks    	  & \xmm\       		& 0743630401 & 2014-09-27 17:30:23 		& 33.5~ks 	  \\
          					& 				  & 					  &                      & \xmm\        		& 0743630501 & 2014-09-28 21:01:46  	         & 39.2~ks 	  \\
\sgra\ w/T2				& 30002002006  	  & 2015-02-25 23:41:07  	  &29.2~ks    	  & \nodata     		& \nodata  		& \nodata       			          & \nodata      \\
\sgra\ w/T2				& 30002002008  	  & 2015-03-31 04:41:07 	  &25.7~ks    	  & \nodata     		& \nodata  		& \nodata                   		& \nodata       \\
\sgra\ w/T2				& 30002002010  	  & 2015-04-01 06:31:07  	  &14.4~ks    	  & \nodata     		& \nodata  		& \nodata                   		& \nodata       \\
\sgra\ w/T2				& 30002002012  	  & 2015-04-02 08:21:07  	  &13.1~ks    	  & \nodata     		& \nodata  		& \nodata                   		& \nodata       
\enddata
\tablecomments{* T1 is CXOGC J174540.0$-$290005, an X-ray transient detected during the observation of the Galactic center magnetar SGR J1745$-$29. **T2 is AXJ 1745.6$-$2901, another X-ray transient going into outburst during the magnetar monitoring, and maintaining in outburst for the following \sgra\ flare observations in 2014 and 2015.}
\label{tab:SgrAobs}
\end{deluxetable*}


\section{Data Reduction}

\subsection{\nustar}

We analyzed all the existing \nustar\ Galactic Center observations with \sgra\ in the FoV, resulting in 27 observations with a total exposure of $\sim1$~Ms.
We reduced the data using the \nustar\ {\it Data Analysis Software NuSTARDAS} v.1.3.1. and HEASOFT v.6.13, filtered for periods of high instrumental background due to SAA passages and known bad detector pixels. 
Photon arrival times were corrected for on-board clock drift and precessed to the Solar System barycenter using the JPL-DE200 ephemeris.
For each observation, we registered the images with the brightest point sources available in individual observations, improving the astrometry to $\sim 4 \asec$.
We used a source extraction region with 50\asec\ radius centered on the radio position of \sgra\ at R.A.=266.41684$^{\circ}$, Decl.=--29.00781$^{\circ}$ (J2000) \citep{Reid2004}.
Then we extracted 3-30~keV light curves in 300 s bins with deadtime, PSF, and vignetting effect corrected.
For all 27 observations we examined the data obtained by both focal plane modules FPMA and FPMB, and made use of those not heavily contaminated by ghost-rays from distant bright X-ray sources. 

To derive the \nustar\ flare spectra, we used the same source region we adopted when extracting the light curves to extract both the source and background spectra.
The source spectrum was extracted from the flaring intervals determined by the flare search method (see Section 4),
The background spectrum was extracted from off-flare intervals for each flare in the same observation. 
Spectra of FPMA and FPMB were combined and then grouped with a minimum of $3\sigma$ signal-to-noise significance per data bin, except the last bin at the high-energy end for which we require a minimum significance of $2\sigma$. 

\subsection{\chandra}

\chandra\ observed \sgra\ 38 times at high spectral resolution with the HETGS during the 2012 XVP campaign \citep{Neilsen2013}. 
Three of these observations were coordinated with the \nustar\ pointings; the details of the overlapping observations are listed in Table  \ref{tab:SgrAobs}. 
For the present analysis, we used the same \chandra\ data extraction as \citet{Neilsen2013}. 
Briefly, this involves processing with standard tools from the {\sc ciao} software package (v.4.5), identifying photons dispersed by the transmission gratings using the diffraction equation, and extracting events from a small extraction region (a 2.5 pixel radius circle for the zeroth order photons and 5 pixel wide rectangular strips for the first order dispersed photons) to limit the background.  
Finally, we extracted 2--8~keV light curves in 300 s bins.

For the spectral analysis, we used the same extraction region as for the light curves to create zeroth order and first order grating spectra and responses.  
Since we are interested in the flares, we extracted spectra for the on-flare and off-flare time intervals separately, using the off-flare periods as background spectra to be subtracted.  
To account for pileup in the zeroth order spectra, we used the pileup kernel developed by \citet{Davis2001}, although the pileup parameter is poorly constrained by the data.

\subsection{\xmm}

We reduced the \xmm\ data using version 13.5.0 of the \xmm\ SAS software.
We extracted the source photons from a circular region with 10\arcsec\ radius centered on \sgra.
For each flare we extracted source photons during the time window defined by the Bayesian block routine, adding 200 s before and after the flare. 
Background photons have been extracted from the same source regions by selecting only quiescent periods. 
The count rate of even the brightest \sgra\ flares are below the pile-up count rate threshold of 2~cts~s$^{-1}$, providing \xmm\ the key advantage of being able to collect pile-up free, and therefore unbiased, spectral information even for the brightest flares.
For more details of the \xmm\ data reduction, see \citet{Ponti2015, Ponti2017}.
 

\section{Flare Search}

\subsection{Flare Search Methods}

For the \nustar\ observations, we applied Bayesian block analysis to the combined FPMA and FPMB light curves as described in \citet{Barriere2014}. 
The Bayesian block analysis addresses the problem of detecting and characterizing local variance in the light curves, e.g. transient phenomena \citep{Scargle2013}. 
This Bayesian statistics based method represents the signal structure as a segmentation of the time interval into blocks (or subintervals) separated by change points.
The statistical properties of the signal change discontinuously at the change points but are constant within one block.
Therefore, the time range of the observation is divided into blocks, where the count rate is modeled as constant within errors.
This analysis has been by far one of the most popular methods for detecting and characterizing \sgra\ X-ray flares \citep{Nowak2012, Neilsen2013, Ponti2015, Mossoux2015}. 

We used the Bayesian block analysis algorithm as described by \citet{Scargle2013}.
The dynamic programming algorithm employs a Monte Carlo derived parametrization of the prior on the number of blocks and finds the optimal location of the change points.
The number of change points is affected by two input parameters: the false positive rate $fpr$, which quantifies the relative frequency with which the algorithm falsely reports detection of change points with no signal present, and the prior estimate of the number of change points, $n_{\rm cp-prior}$.
For the \nustar\ data, we adopted the same parameters as used in \citet{Barriere2014}, i.e. $fpr=0.01$ and a geometric prior $n_{\rm cp-prior}=4-\log(fpr/0.0136~N^{0.478})$, where $N$ is the total number of events.

The same Bayesian block analysis algorithm was modified to read \xmm\ events files and applied to all the \xmm\ observations as well, as described in \citet{Ponti2015}.
For the \chandra\ observations, both direct fits (with one or more Gaussian components superimposed on a constant background) and Bayesian block analysis were adopted for the \chandra\ X-ray light curves to detect and characterize X-ray flares, as described in detail in \citet{Neilsen2013} and \citet{Ponti2015}.
The properties of the detected \chandra\ flares are not sensitive to the detection algorithm. 

\subsection{Flare Detection Results}

As our \nustar\ X-ray flare database gets larger, from now on we name all the flares in chronological order, along with other publication names, if any.
Table \ref{tab:FlareDet} lists the name, start time, duration and detection significance for the ten flares as detected by \nustar, and as detected by \chandra\ or \xmm\ if there is a simultaneous observation. 

\subsubsection{2012 Joint \sgra\ Observing Campaign and Mini-survey: Six Flares Detected}
  
For the three 2012 \nustar\ \sgra\ observations (ObsID 30001002001, 30001002003, 30001002004), the Bayesian block analysis led to detection of four bright X-ray flares from \sgra\ (for details see \citealp{Barriere2014}).
Three out of the four bright flares were detected in a row within $\sim20$~hrs from 2012 July 20 to July 21, named as flares Nu1(J20), Nu2(J21-1) and Nu3(J21-2) with durations of $\sim920$~s, $\sim1238$~s and $\sim3099$~s respectively.
The baseline count rate of the \sgra\ region in 3--79~keV is $0.59\pm0.01$~cts~s$^{-1}$ (all count rates given with 1$\sigma$ error bars).
The baseline emission is dominated by faint X-ray point sources and diffuse emission around \sgra, while the instrument background contributes  $<5\times10^{-3}\rm~cts~s^{-1}$. 
During the flares, the count rate in the same source region reaches $0.73\pm0.03\rm~cts~s^{-1}$ for flare J20, $0.80\pm0.03\rm~cts~s^{-1}$ for Nu2(J21-1), and $1.05\pm0.02\rm~cts~s^{-1}$ for Nu3(J21-2).
The fourth bright flare, noted as Nu6(O17), reported in \citet{Barriere2014} was simultaneously detected by \chandra\ and \nustar\ on 2012 October 17. 
This bright flare results in a significant detection level of $\ge10\sigma$ for both X-ray observatories.
Compared with the full profile of this flare obtained by \chandra, \nustar\ captured the peak $\sim1249$~s of the flare.
The \nustar\ flare peak count rate reaches $1.20\pm0.02\rm~cts~s^{-1}$, while the baseline emission maintains at the same level as in the 2012 July observation ($0.59\pm0.01$~cts~s$^{-1}$).

Below we report two new flares detected from the 2012 Galactic Center observation campaign.
First, to search for fainter flares, we compared the \nustar\ observations with the simultaneous \chandra\ observations. 
In the coordinated 2012 \chandra\ observations (ObsID 13842, 13852, 13851), the direct fit algorithm detected seven flares, which was further confirmed by the Bayesian block analysis method (Table 1, \citealp{Neilsen2013}).
By comparing the duration of these seven \chandra\ flares and the \nustar\ observation good time intervals (GTIs), we found two more flares covered by the \nustar\ observations. 
For one of the two flares, merely $\sim100$~s of exposure time is covered by the \nustar\ GTIs, resulting in poor statistics for any meaningful analysis.
We therefore exclude this flare from our study. 
The other faint flare was detected by \chandra\ on 2012 August 5 with a $\sim3\sigma$ detection.  
The \nustar\ GTIs of the observation 30001002003 partly covered this flare, resulting in a marginal detection ($\sim2.5\sigma$).
While the \sgra\ region baseline emission remains the same as in 2012 July ($0.59\pm0.01$~cts~s$^{-1}$), the \nustar\ 3--79~keV count rate of this flare is $0.64\pm0.02$~cts~s$^{-1}$.
Because of its low count rate relative to the baseline count rate, flare Nu4 is not significant in the \nustar\ data alone.
 
We also searched for \sgra\ flaring activities using the observations from 2012 \nustar\ Galactic Center Mini-survey \citep{Mori2015}.
Four of the six observations have the \sgra\ region included in the FoV (ObsID 40010001002, 40010002001, 40010003001, 40010004001).
We performed the Bayesian block analysis on these four observations, following the procedures described in \citet{Barriere2014}.
An increase of \sgra\ X-ray flux is detected at $\sim3.3\sigma$ significance level on 2012 October 15 (hereafter flare Nu5).
During 2012 October, the \sgra\ baseline emission count rate is $0.57\pm0.01\rm~cts~s^{-1}$, consistent with that of 2012 July, while the count rate during flare Nu5 is $0.80\pm0.07$~cts~s$^{-1}$. 
There were no joint observations of the Galactic Center during the Mini-Survey, so we have no additional constraints on the properties of the flare.

\subsubsection{2013 \nustar\ Galactic Center Transient Observations: No Flares Detected}
When the magnetar SGR~J1745-29 (merely 2.4\asec\ away from \sgra) went into outburst in 2013 April with a peak flux of $F_{1-10\rm~keV} \sim 2\times10^{-11}\rm~erg~cm^{-2}~s^{-1}$, the \sgra\ source region was dominated by the X-ray emission from the magnetar (e.g. \citealp{Mori2013, Rea2013, Ponti2015}). 
The severe contamination from the magnetar prevents a clear detection and clarification of even bright X-ray flares for observations 30001002006 to 80002013024 (see Table \ref{tab:SgrAobs}).
During the magnetar monitoring campaign, flare detections further suffered from PSF wing contamination from two nearby X-ray transients CXOGC~J174540.0$-$290005 and AXJ~1745.6$-$2901, which went into outburst in 2013 May and July respectively (see Section 2).
The baseline emission from the \sgra\ area was therefore highly variable due to contamination from the three bright X-ray transients.
A routine flare search via Bayesian block analysis on the $\sim380$~ks Galactic Center observations conducted in 2013 found no significant \sgra\ flaring activity, as \nustar\ was not sensitive to flares with luminosities lower than 50 times the \sgra\ quiescent luminosity during this period.

\subsubsection{2014 Joint Observing Campaign: \\ No X-ray Flares Detected}
During the 100~ks \sgra\ observations coordinated with \chandra\ and \spitzer (obsID 30001002008, 30001002010 for \nustar; obsID 16597 for \chandra)
the X-ray flux of the magnetar SGR~J1745-29 had dropped to $F_{1-10\rm~keV} \sim2\times10^{-12}\rm~erg~cm^{-2}~s^{-1}$, allowing adequate characterization of \sgra\ X-ray flares.
In the 16.5~ks \chandra\ observation (obsID 16597), we found no \sgra\ flaring activity via a direct light curve fit.
Since the X-ray transient AX~J1745.6$-$2901 was still bright in our observation, it increased the \nustar\ baseline count rate to $0.84\pm0.02$~cts~s$^{-1}$, which is $\sim50\%$ higher than in 2012.
Due to the increased baseline emission from the transient, we can only say that there were no flares with luminosities above 20 times the quiescent luminosity during this campaign.
Around 2014 June 18 UT 09:24, \sgra\ flaring activities were detected by \spitzer, but we found no X-ray counterpart for this flare.
The \spitzer\ flare characteristics will be discussed elsewhere.

\subsubsection{2014-2015 Joint Observing Campaign: \\ Four Flares Detected}
Four X-ray flares were simultaneously detected by \xmm\ and \nustar\ in 2014 fall (obsID 30002002002, 30002002004 for \nustar; obsID 0743630201, 0743630301, 0743630401, 0743630501 for \xmm).
Three out of the four flares were detected in a row within $\sim26$~hrs on 2014 August 30, August 31 and September 1, hereafter flare Nu7, Nu8 and Nu9.
\xmm\ was able to capture the full flare profile for all three flares \citep{Ponti2015, Ponti2017}. 
However, due to interruptions caused by Earth occultations, \nustar\ GTIs only captured the rising half (1215 s) of flare Nu7, 518 s of the rising stage of flare Nu8, and half of flare Nu9 (see Figure 1).

This is the second time that multiple flares are detected by \nustar\ roughly within one day, which could suggest that bright flares tend to take place in clusters, as also indicated by previous flare studies \citep{Porquet2008, Ponti2015}.
The transient source AXJ~1745.6$-$2901 continued to stay in outburst, therefore continuing to contaminate the \sgra\ region.
During the 2014 Fall \nustar\ observation, the baseline emission from the \sgra\ region was $0.78\pm0.02$~cts~s$^{-1}$, about 30\% higher than that in 2012. 
\xmm\ also detected a fainter X-ray flare on 2014 September 29.
The \nustar\ observation in the same time range results in a $2\sigma$ detection (hereafter Nu10).

\begin{deluxetable*}{lcrrcccr}   
\centering                                                                                                    
\tablecaption{\nustar\ Flares and simultaneous detection by \chandra/\xmm}
\tablewidth{0pt}
\tablecolumns{7}                                                                                                                    
\tablehead{ \colhead{\nustar}   & \colhead{ }   & \colhead{ }  & \colhead{ }  &  \colhead{Joint Obs}  &  \colhead{ }  &\colhead{ } &\colhead{ }  \\
\colhead{Flare}  &\colhead{Start (UT)}  &\colhead{Coverage(s)} &\colhead{Significance($\sigma$)} &\colhead{Instrument}   &\colhead{Start(UT)}  &\colhead{Duration(s)}   &\colhead{Significance($\sigma$)}  }
\startdata
Nu1 (J20)		     & 2012-07-20 12:15:21 		& 920   		& 5 		& \nodata 		& \nodata 					& \nodata	 	& \nodata 		\\
Nu2 (J21-1)	     & 2012-07-21 01:45:15 		& 1238 		& 7 		& \nodata 		& \nodata 					& \nodata 		& \nodata          \\
Nu3 (J21-2)	     & 2012-07-21 06:01:12 		& 3099 		& 20 		& \nodata 		& \nodata  				&\nodata  		& \nodata          \\
Nu4               	     & 2012-08-05 08:20:17  		& 1319  		& 2   		& \chandra\ 	& 2012-08-05 07:41:54 		& 3623 		& 3                    \\   
Nu5               	     & 2012-10-15 01:11:10 		& 822  		& 3  		& \nodata 		& \nodata 					& \nodata 		& \nodata          \\
Nu6 (O17)             & 2012-10-17 19:50:08 		& 1249 		& 20 		& \chandra\ 	& 2012-10-17 19:35:09 		& 5900 		& 11                   \\
Nu7 (VB3)     	     & 2014-08-30 23:44:15 		& 1215 		& 14 		& {\it XMM}       & 2014-08-30 23:42:08 		& 2727 		& 10                   \\
Nu8  (B3) 		     & 2014-08-31 04:23:41 		& 1104 		& 8   		& {\it XMM}       & 2014-08-31 04:31:35 		& 1469 		& 6                     \\
Nu9  (B4) 		     & 2014-09-01 01:08:17 		& 2175 		& 5   		& {\it XMM}       & 2014-09-01 00:43:38 		& 4359 		& 15                   \\
Nu10 (B5) 	     & 2014-09-29 06:06:55 		& 6273 		& 2  		& {\it XMM}       & 2014-09-29 06:06:55 		& 7655 		& ~~6                     
\enddata
\tablecomments{The flare names are given in chronological order (along with other publication names, if any). 
Flares Nu1(J20), Nu2(J21-1), Nu3(J21-2) and Nu6(O17) were previously reported in \citet{Barriere2014}. The \chandra\ data of flare Nu4 is discussed in \citet{Neilsen2013}. 
The multi-wavelength observation of the flares Nu7(VB3), Nu8(B3), Nu9(B4) and Nu10(B5) are reported in \citet{Ponti2015, Ponti2017}. }
\label{tab:FlareDet}
\end{deluxetable*}

\begin{figure*}
\centering
\label{fig:FlareLC}
\begin{tabular}{cc}
\includegraphics[width=0.45\linewidth]{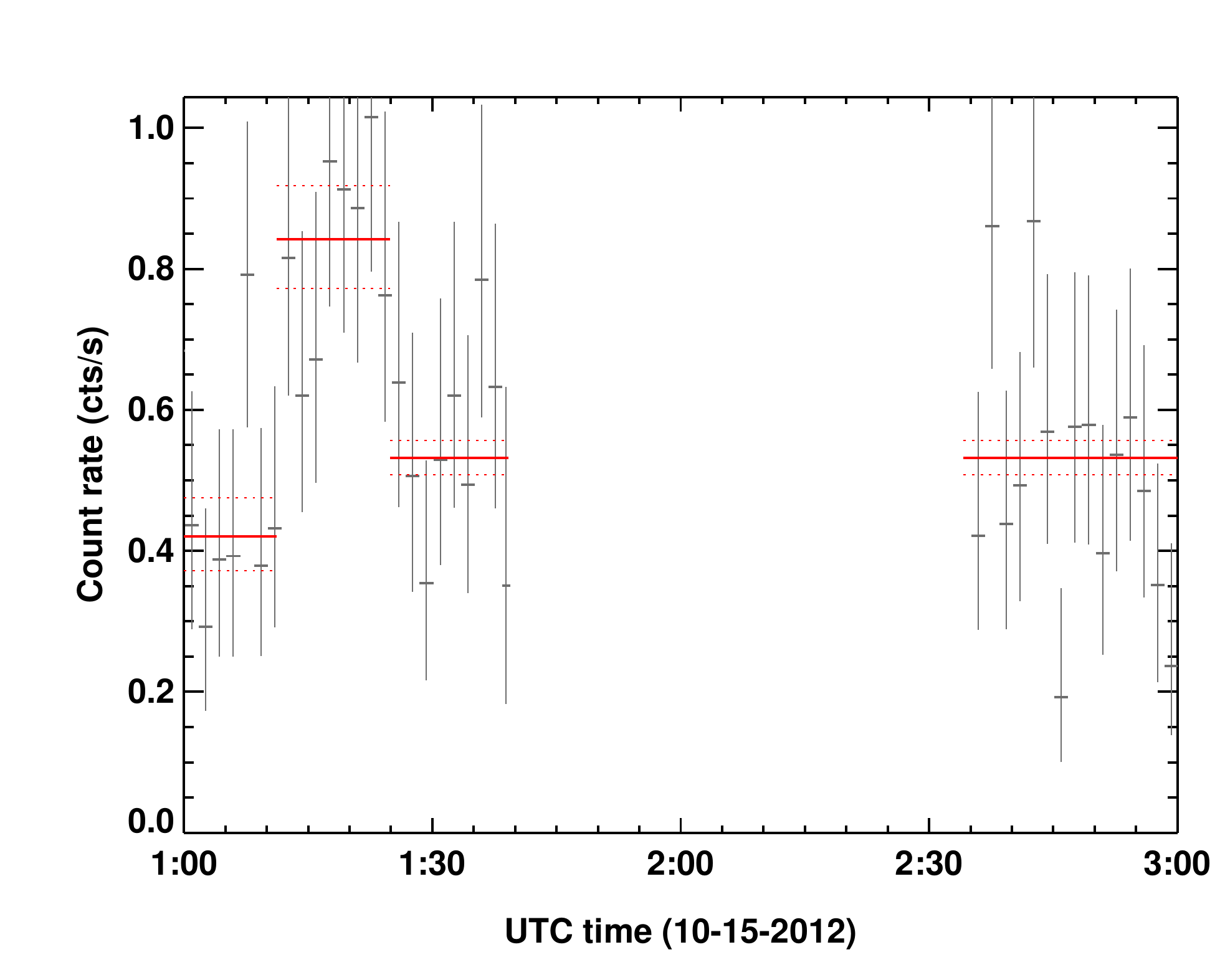} &
\includegraphics[width=0.45\linewidth]{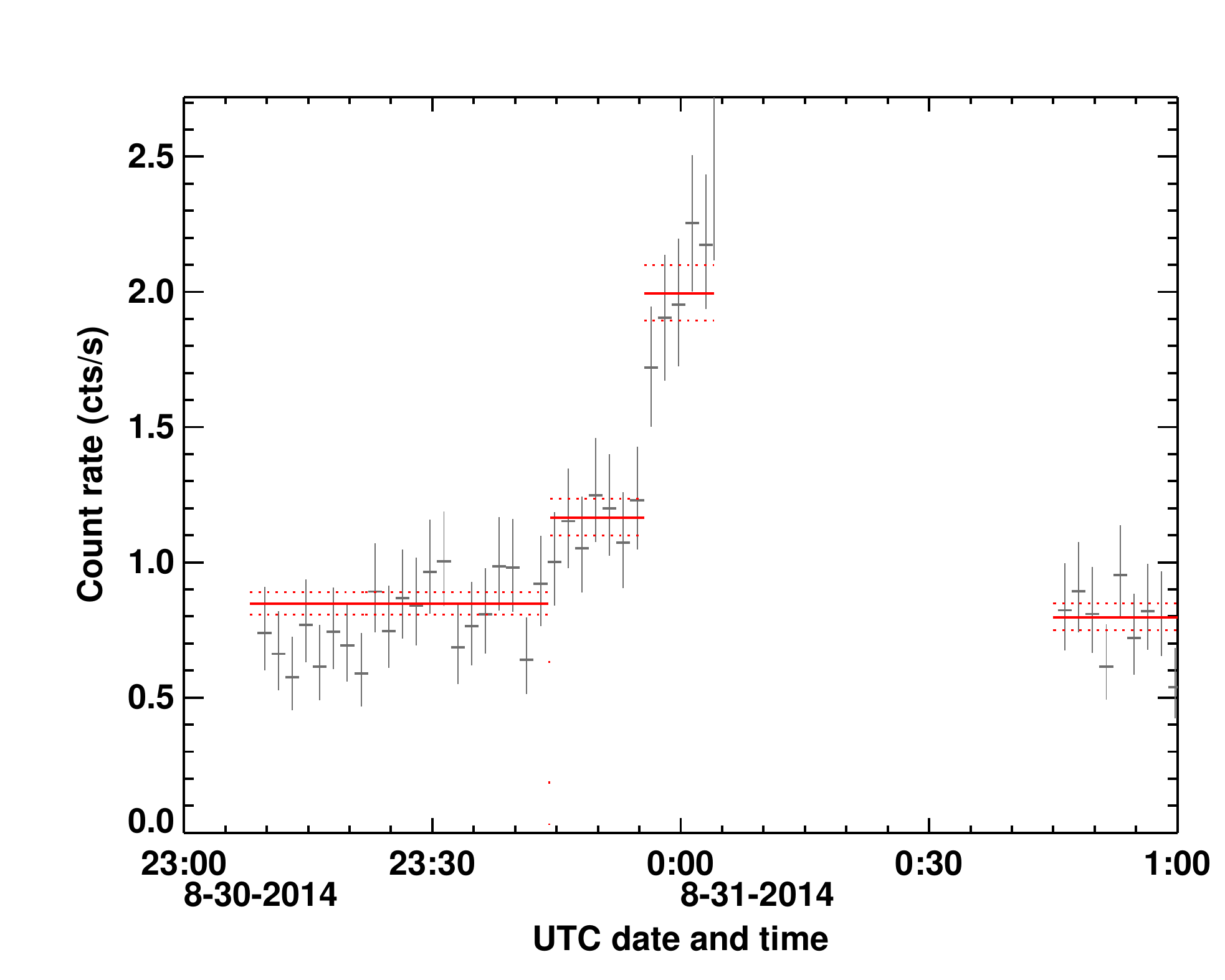} \\
\includegraphics[width=0.45\linewidth]{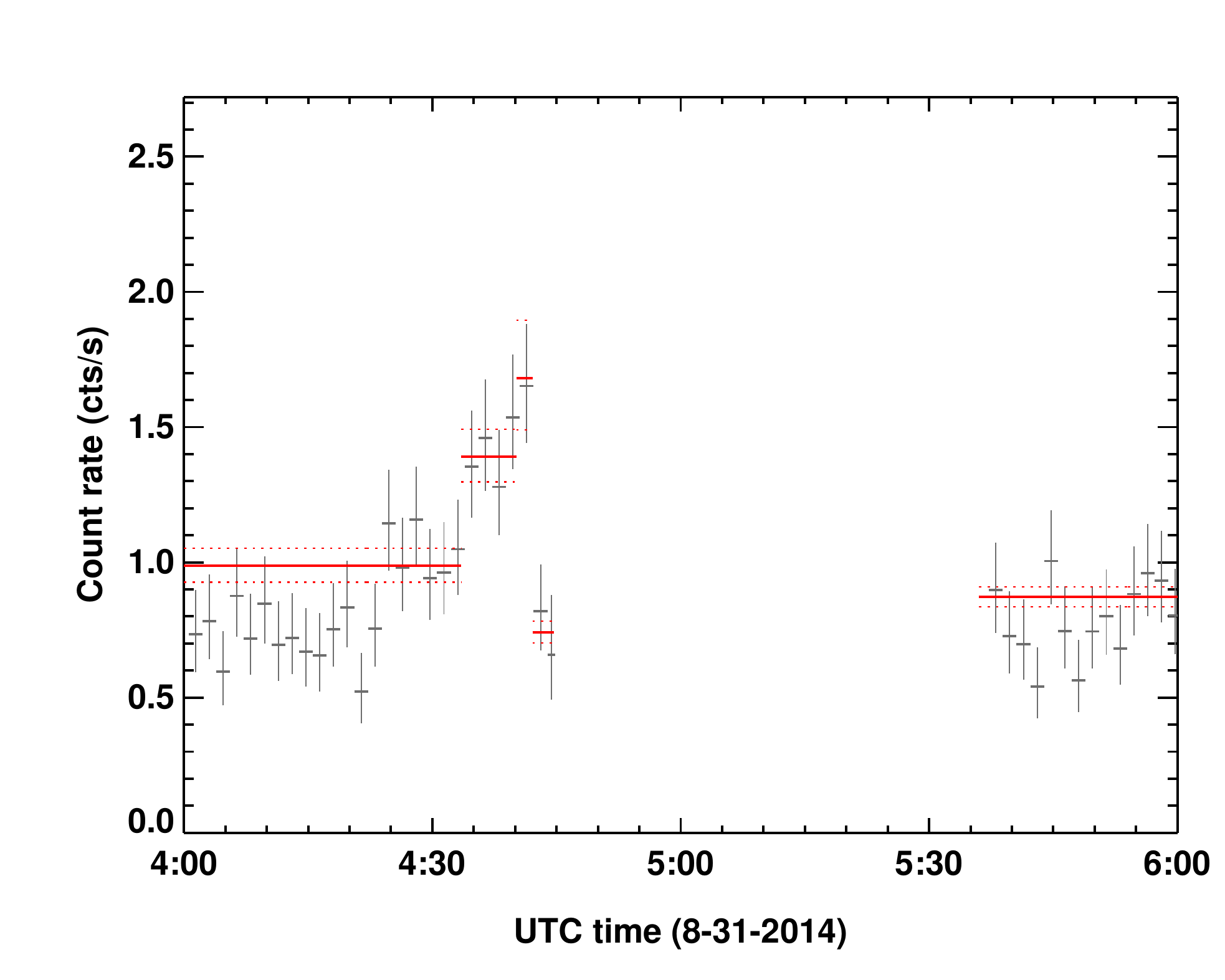} &
\includegraphics[width=0.45\linewidth]{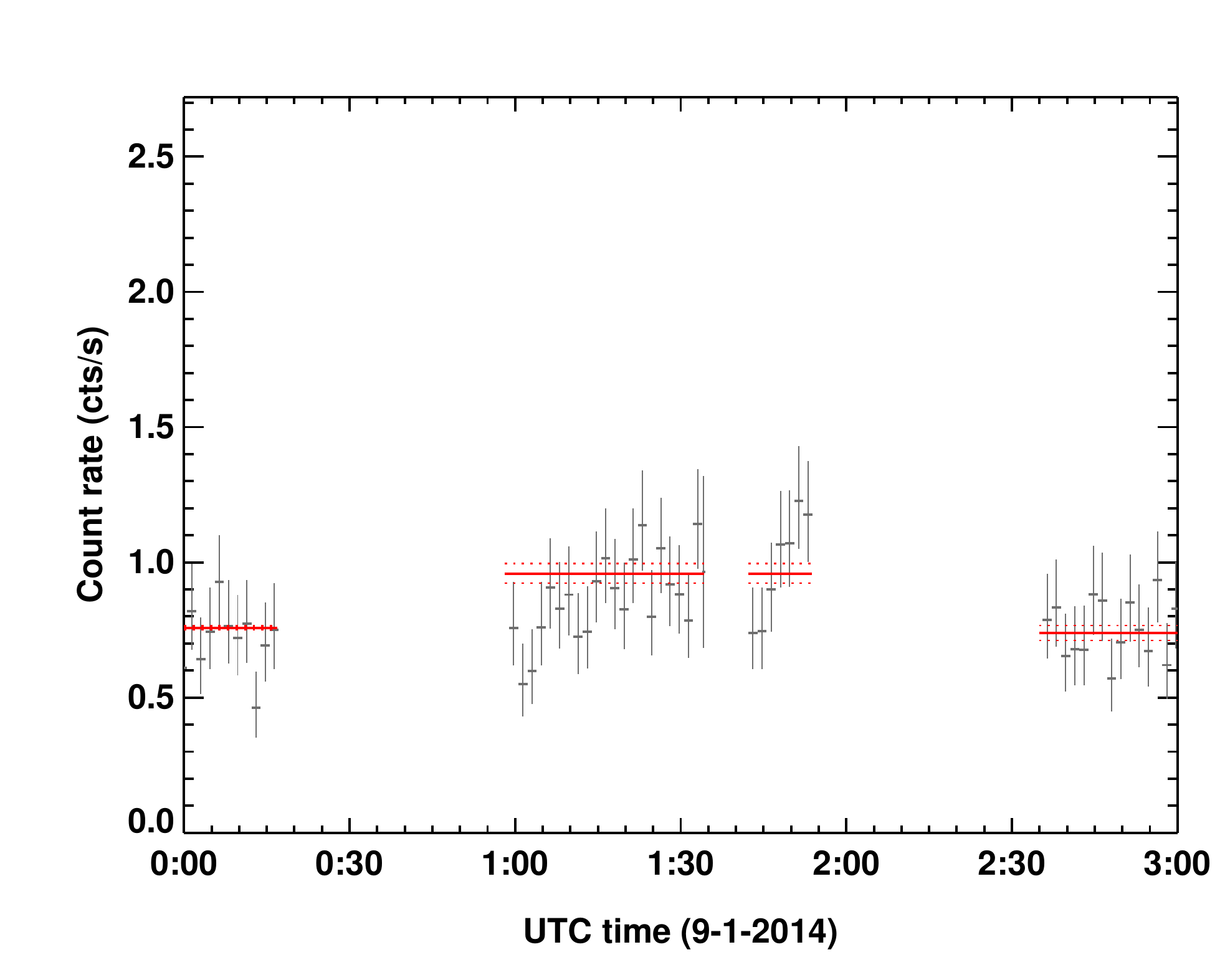} \\
\end{tabular}
\caption{\nustar\ $3-79$~keV light curves showing previously unreported flares with $>3\sigma$ detections, including flare Nu5 (upper left), Nu7 (upper right), Nu8 (lower left) and Nu9 (lower right). The \nustar\ light curves are deadtime, PSF, and vignetting corrected and extracted from a 30\asec\ radius circle centered on \sgra\ in 100s bin. The light curves of the four bright flares Nu1, Nu2, Nu3 and Nu6 are shown in Figure 1 and 2 in \citet{Barriere2014}. Flares Nu4 and Nu10 are not significantly detected with \nustar\ data only. The Nu4 \chandra\ light curve is presented in \citet{Neilsen2013}; the Nu10 \xmm\ light curve is presented in \citet{Ponti2015}.}
\end{figure*}


\section{Flare Spectral Properties}

\subsection{The Brightest X-ray Flare Detected by \nustar}

Flare Nu6 is the brightest X-ray flare detected by \nustar.
It was simultaneously detected by both \nustar\ and \chandra.
While \chandra\ captured the full flare lasting $\sim5900$~s \citep{Neilsen2013},
\nustar\ only captured the peak $\sim1249$~s of the flare, mainly due to interruption by Earth occultation.

The \chandra\ data does not show spectral evolution within this flare, so we jointly fitted the 1249~s \nustar\ flare peak spectrum in 3--79~keV and the $\sim5900$~s \chandra\ full flare spectrum in 0.5--9~keV.  
We used the Interactive Spectral Interpretation System v.1.6.2-19 \citep{HouckDenicola2000}, setting 
the atomic cross sections to \citet{Verner1996} and the abundances to \citet{Wilms2000}. 
The joint spectrum is well-fit by a simple absorbed power-law, with the dust scattering 
taken into account for the \chandra\ spectra ({\tt Tbabs*dustscat*powerlaw}) \citep{Baganoff2003, Neilsen2013}.   
We did not use the dust scattering model for the \nustar\ spectra, 
because with a large extraction region, the photons scattering into and out of the line of sight compensate with each other \citep{Barriere2014}.
The best-fitted photon index is $\Gamma=2.06^{+0.19}_{-0.16}$ with an absorption column 
density $N_{\rm H} = (1.5^{+0.3}_{-0.2}) \times10^{23}~{\rm cm}^{-2}$ (Table \ref{tab:specfit}).
Both the photon index and the column density are consistent with those derived from \nustar\ spectrum alone ($\Gamma=2.04^{+0.22}_{-0.20}$, $N_{\rm H}=(1.7^{+0.7}_{-0.6}) \times 10^{23}\rm~cm^{-2}$, \citealp{Barriere2014}), though better constrained. 
The spectrum with the best-fit absorbed power-law model for the flare is shown in Figure 2. 
The 0.5--79~keV unabsorbed flare peak flux is $F_{X}=(6.2 \pm 0.6)\times10^{-11}$~erg~cm$^{-2}$~s$^{-1}$, corresponding to a luminosity of 
$L_{X}=(4.7 \pm 0.5) \times 10^{35}$~erg~s$^{-1}$ assuming the distance to the Galactic Center is 8.0 kpc \citep{Reid2004}.
This is by far the brightest X-ray flare detected by \nustar\ and one of the brightest flares detected by \chandra.

\begin{figure}
\label{fig:O17spectra}
\includegraphics[width=1.0\linewidth]{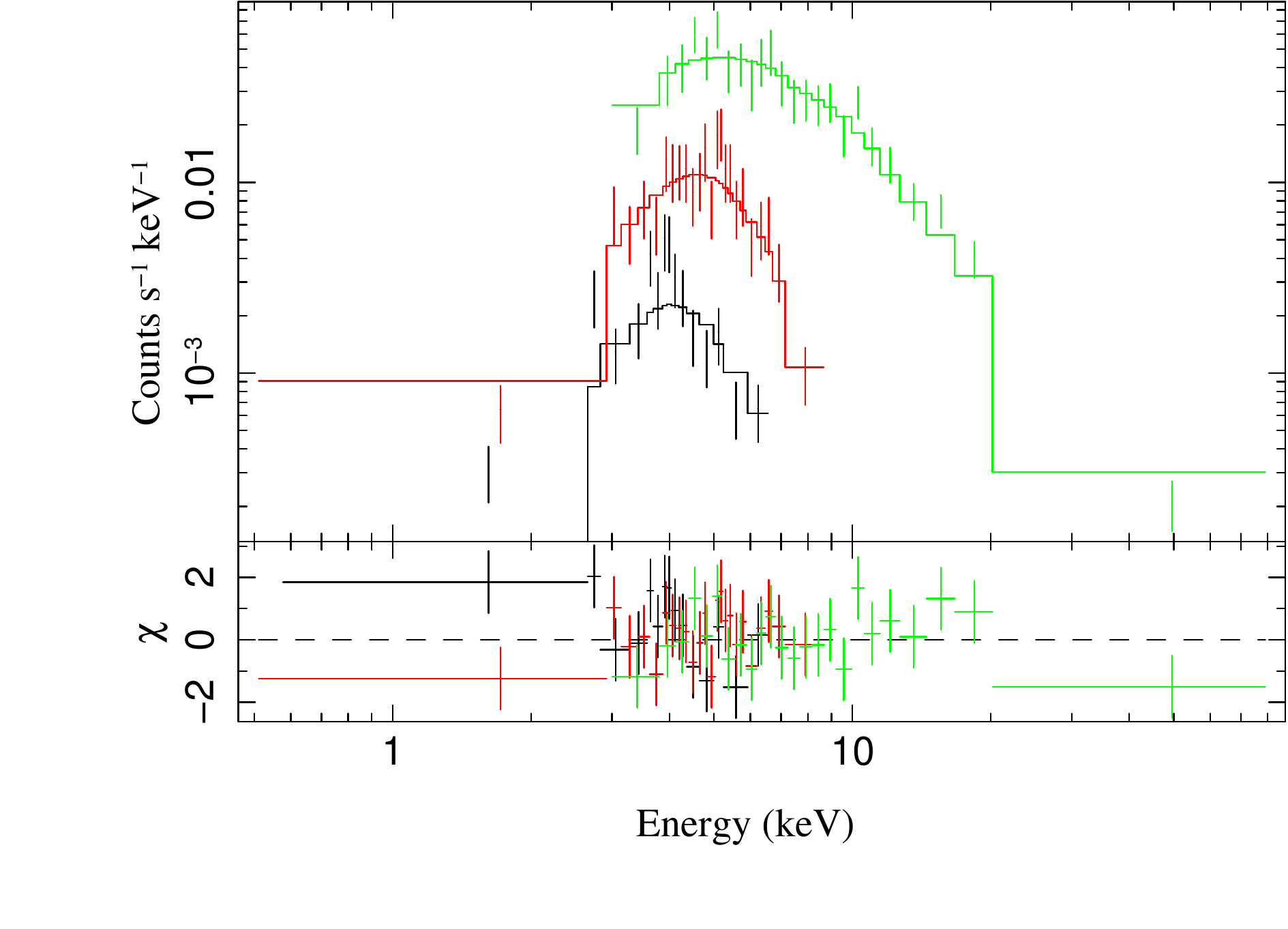}
\caption{\nustar\ FPMA and FPMB combined spectra (green) and  \chandra\ zeroth-order and $1^{st}$ order spectra (black and red, respectively) for flare Nu6 peak jointly fitted to an absorbed power-law model. 
The crosses show the data points with 1-$\sigma$ error bars, and the solid lines show the best fit model. The lower 
panel shows the deviation from the model in units of standard deviation.}
\end{figure}

\begin{deluxetable}{ll}[H]                                                                                                        
\tablecaption{Power-law model for the {\it Chandra} and {\it NuSTAR} data of flare Nu6. }
\tablecolumns{2}                                                                                                                    
\tablehead{ \colhead{Parameters}   &  \colhead{Value}  }
\startdata
$N_{\rm H}$ (10$^{23}$~cm$^{-2}$)             & $1.5^{+0.3}_{-0.2}$ \\
$\Gamma$                                                    &  $2.06^{+0.19}_{-0.16}$ \\
Flux ($10^{-11}$~erg~cm$^2$~s$^{-1}$)     & $6.2 \pm 0.6$ \\
$\chi^2_{\rm \nu}$ (DoF)                              & 0.94 (57) \\
\enddata
\tablecomments{$N_{\rm H}$ is the column density,  $\Gamma$ is the photon index of the power-law. 
The unabsorbed flux is given in 0.5--79 keV. The goodness of fit is evaluated by the reduced $\chi^2$ and the degrees of freedom is given in parentheses. 
The errors are at 90\% confidence level.}
\label{tab:specfit} 
\end{deluxetable}


\subsection{Spectral Properties of All Ten Flares}

We analyzed the X-ray spectra of all the X-ray flares detected by \nustar, jointly with either \chandra\ or \xmm\ when available.
We extracted the source spectra from the flare time ranges, and the background spectra from off-flare time ranges.
We first focused on the seven flares that are detected with $>5\sigma$ detection significance, i.e. flares Nu1, Nu2, Nu3, Nu6, Nu7, Nu8 and Nu9.
The first set of four flares (Nu1, Nu2, Nu3, Nu6) were detected in 2012 Fall, when no X-ray transient in the Galactic Center was detected.
Among them, flare Nu6 was simultaneously detected by \chandra.
The second set of three flares (Nu7, Nu8, Nu9) was detected jointly by \nustar\ and \xmm\ in 2014 fall, during which AX~J1745.6$-$2901 was still in outburst and increased the \sgra\ off-flare baseline emission by $\sim30\%$ through PSF contamination (see section 4.2.4).
Therefore, varying baseline emission is an aspect of our data set.
In order to make a fair comparison of the flare spectral shapes, below we first examined two factors that could affect joint fitting of all seven flares: 1) AX~J1745.6$-$2901 PSF contamination; 2) absorption column density.

First, we checked how the contribution from the transient AX~J1745.6$-$2901 would affect the measurements of the 2014 flares Nu7, Nu8 and Nu9.
We investigated the light curve and the spectrum of the transient AXJ~1745.6$-$2901 during the flare and the off-flare time ranges in the 2014 observation (obsID: 30002002002) where the second set of three bright flares were detected.
Throughout this observation, the transient does not demonstrate significant variation except for eclipses. 
The 3--79~keV count rate in the 30\asec\ region centered on AX~J1745.6$-$2901 maintains at $2.00\pm0.02$~cts~s$^{-1}$, while during the eclipse the count rate dropped to $0.34\pm0.02$~cts~s$^{-1}$.
No eclipse coincides with any of the three flares.
Therefore, when selecting background spectra during the off-flare time range, we excluded eclipses.
Next, we compared the spectra of AX~J1745.6$-$2901 during and off the flares.
Both can be well fit with a simple absorbed power-law model, yielding $N_{\rm H}=(1.8\pm0.2) \times10^{23}$~cm$^{-2}$ and $\Gamma=1.77\pm0.03$ with an absorbed 3--79~keV flux of $F_{3-79\rm~keV}\sim9.5\times10^{-11}$~erg~cm$^{-2}$~s$^{-1}$.
The absorbed 3--79~keV flux during and off the flares was constant at $F_{3-79\rm~keV}\sim9.5\times10^{-11}$~erg~cm$^{-2}$~s$^{-1}$.
Therefore, the PSF contamination from AX~J1745.6$-$2901 within the \sgra\ region does not have significant variation during and off the flares, and thus can be treated as a constant contribution to the baseline spectrum.
However, this elevated off-flare baseline emission from the \sgra\ region in 2014 (30\% higher than in 2012) does cause larger error bars for spectral properties of flares Nu7, Nu8 and Nu9.

Second, we investigated whether the absorption column density $N_{\rm H}$ varies from 2012 to 2014.
We fit the two sets of \nustar\ flare spectra separately with an absorbed power-law model, and found that the best-fit values of the absorption column density for each set are consistent with each other, resulting in $N_{\rm H}=(1.7^{+0.7}_{-0.6})\times10^{23}~$cm$^{-2}$ for the first set of spectra and $N_{\rm H}=(1.7^{+0.9}_{-0.8})\times10^{23}\rm~cm^{-2}$ for the second set of spectra.
Therefore, here we can safely assume that the absorption column density did not vary with time (see \citealt{Ponti2017, Jin2017} for more details).

After investigating the above factors, we proceeded to joint spectral fitting of the seven bright flares. 
We use the same model as described in Section 5.1, i.e. {\tt Tbabs*powerlaw} for the \nustar\ flare spectra and 
{\tt Tbabs*dustscat*powerlaw} for the \chandra\ and \xmm\ flare spectra. 
The absorption column density values $N_{\rm H}$ are tied among all the spectra.
The photon indices of the spectra associated with the same flare obtained by different instruments are tied; the photon indices of different flares are independent.
The power-law normalization are set free.
We then performed a joint fit of the seven bright X-ray flares using all available data, resulting in a good fit with $\chi_{\nu}^{2}=1.02$ with DoF of 295.
The resultant column density is $N_{\rm H}=(1.6\pm0.2) \times10^{23}~$cm$^{-2}$.
Table \ref{tab:FlareSpec} lists the corresponding best-fit photon index, flux and luminosity for each flare.
We also calculated the strength $S$ of each flare, which is defined as the ratio of the unabsorbed 2--10~keV flare flux and the quiescent state when $F_{q}=(0.47^{+0.04}_{-0.03})\times10^{-12}$~erg~cm$^{-2}$~s$^{-1}$ \citep{Nowak2012}.

Faint flares with strengths less than 30 times the \sgra\ quiescent flux (Nu1, Nu2 and Nu9, black in Figure 3) have best-fit photon indexes of $\Gamma=(2.2$--$2.8)\pm(0.6-1.0)$; flares with strengths higher than 30 times while lower than 50 times the \sgra\ quiescent flux (Nu3, Nu7 and Nu8, red in Figure 3) have best-fit photon indexes of $\Gamma \sim2.3\pm(0.2$--$0.5)$.
The brightest flare, Nu6, with strength $\sim54$ times the \sgra\ quiescent flux, has the hardest spectrum with a photon index of $\Gamma=2.06\pm0.17$ (green in Figure 3).
To investigate whether brighter flares possess harder spectra, we performed a linear fit to the flare photon index over their strength.
We found that the data can be best fit with a linear function $\Gamma=(-0.016\pm0.010) \times \Gamma+(2.9\pm0.5)$, with a slope of $a=-0.016\pm0.010$ and $\Gamma_{0}=2.9\pm0.5$ (the error bars are given in $1\sigma$ significance level).
Given the low significance ($<2\sigma$)of the slope, our results is consistent with no hardening spectra for brighter flares.
While the best-fit spectral hardening is $\Delta \Gamma=-0.6$ for flares with strengths from $S=18$ to $S=54$, a spectral hardening of $|\Delta \Gamma| > 1.7$ and a spectral softening of $|\Delta \Gamma| > 0.5$ can be excluded.
A Spearman rank correlation test results in $P > 0.10$ (with Spearman's $\rho$ of $-0.36$), confirming that no strong correlation has been found. 
Therefore, our current flare dataset does not show an obvious correlation between flare spectral shape and luminosity, although such dependence cannot be excluded.
This result is consistent with previous works \citep{Porquet2008, Nowak2012, Neilsen2013, Degenaar2013, Ponti2017}.

As we now have a larger sample of flares detected in a broad X-ray energy band, we investigated whether the flare spectra require any curvature or spectral breaks by accumulating all the flares.
We fit the seven flares with the same absorbed power-law model and parameter settings as discussed above, except that the photon indices $\Gamma$ of all the data sets are now tied with each other.
This results in an equally good fit, with $\chi_{\nu}^{2}=1.01$ for DoF of 301.
We derived a best fit column density of $N_{\rm H}=(1.5\pm0.2)\times 10^{23}$~cm$^{-2}$ and the flare photon index of $\Gamma=2.2\pm0.1$.
A spectral break is not required by this data set.
An energy break below 20~keV can be ruled out by the data.
We thus conclude that there is no evidence for a spectral break with this larger flare spectrum sample.  

For the three flares with detection significance lower than $5\sigma$ (due to low luminosity or limited time coverage), we tried a joint fitting with absorbed power-law models using Cash statistics \citep{Cash1979}.
While fixing $N_{\rm H}$ to $1.5\times10^{23}~$cm$^{-2}$, the photon indices of the three flares cannot be well constrained, resulting in $\Gamma=(2-3)\pm3$.
All three flares possess luminosity less than 20 times the quiescent level.

\begin{deluxetable*}{lccccc}                                                                                                       
\tablecaption{\nustar\ Flares and simultaneous detection by \chandra/\xmm}
\tablewidth{0pt}
\tablecolumns{7}                                                                                                                    
\tablehead{ \colhead{Flare}     & \colhead{$\Gamma$}  & \colhead{$F_{abs, 3-79\rm~keV}$}                                  & \colhead{$L_{3-79\rm~keV}$}       & \colhead{Flare strength}\\
                    \colhead{ }         &\colhead{ }                      &\colhead{($10^{-11}$~erg~cm$^{-2}$~s$^{-1}$)}           &\colhead{($10^{35}$~erg~s$^{-1}$)}   & \colhead{}   }
\startdata
Nu1 (J20)       	& $2.6\pm0.9$		                 & $0.7^{+0.6}_{-0.3}$  	& $0.7^{+0.4}_{-0.3}$ 	         & $18^{+13}_{-8}$	\\
Nu2 (J21-1)  	& $2.8\pm0.6$				& $0.9\pm0.3$ 			& $0.9\pm0.3$  			 & $25^{+13}_{-8}$	\\
Nu3 (J21-2)	& $2.3\pm0.2$		                 & $2.2\pm0.4$ 			& $2.1\pm0.3$  			 & $35^{+10}_{-7}$	\\
Nu6 (O17)   	& $2.1\pm0.2$				& $4.4\pm0.7$  		& $4.0\pm0.5$  			 & $54^{+14}_{-11}$ 	\\
Nu7 (VB3)   	& $2.3\pm0.2$				& $2.4\pm0.3$  		& $2.3\pm0.3$  			 & $43^{+11}_{-9}$	\\
Nu8 (B3)     	& $2.2\pm0.5$				& $1.8\pm0.3$ 			& $1.7\pm0.3$  			 & $34^{+10}_{-9}$ 	\\
Nu9 (B4)     	& $2.2\pm0.6$			     	& $0.8^{+0.5}_{-0.3}$ 	& $0.7^{+0.4}_{-0.2}$   	 	 & $15^{+11}_{-7}$ 	\\
\hline
Nu5     		& $3\pm2$				& $0.5\pm0.3$	   		& $0.6\pm0.3$  			 & $18^{+11}_{-9}$	\\
Nu4           	& $2\pm2$	 			& $0.4^{+0.3}_{-0.2}$ 	& $0.3^{+0.3}_{-0.2}$ 	 	& $4^{+4}_{-3}$ 		\\
Nu10 (B5)   	& $3\pm3$ 				& $\le 0.15$ 			& $\le0.2$  				 & $\le6$
\enddata
\tablecomments{The second column gives the best-fit photon index $\Gamma$. The fluxes are determined using the {\tt cflux} convolution model.
The column density of $N_{\rm H}=(1.55^{+0.21}_{-0.19})\times10^{23}\rm~cm^{-2}$ is determined by jointly fitting the seven bright X-ray flares with the $N_{H}$ tied together.
For the three flares detected at low significance (in the lower part of the table), the column density is fixed to $N_{\rm H}=1.55\times10^{23}\rm~cm^{-2}$.
Absorbed flux (noted as $F_{abs}$) and corresponding luminosity assumes a distance of 8~kpc with isotropic emission.
The strength is defined as the ratio of the 2--10~keV unabsorbed flare flux to the 2--10~keV unabsorbed \sgra\ quiescent flux of $F_{q}=(0.47^{+0.05}_{-0.03})\times10^{-12}\rm~erg~cm^{-2}~s^{-1}$ \citep{Nowak2012}.
All uncertainties are reported at the 90\% confidence level. 
}
\label{tab:FlareSpec}
\end{deluxetable*}

\begin{figure*}
\centering
\label{fig:FlareIvsS}
\includegraphics[width=0.6\linewidth]{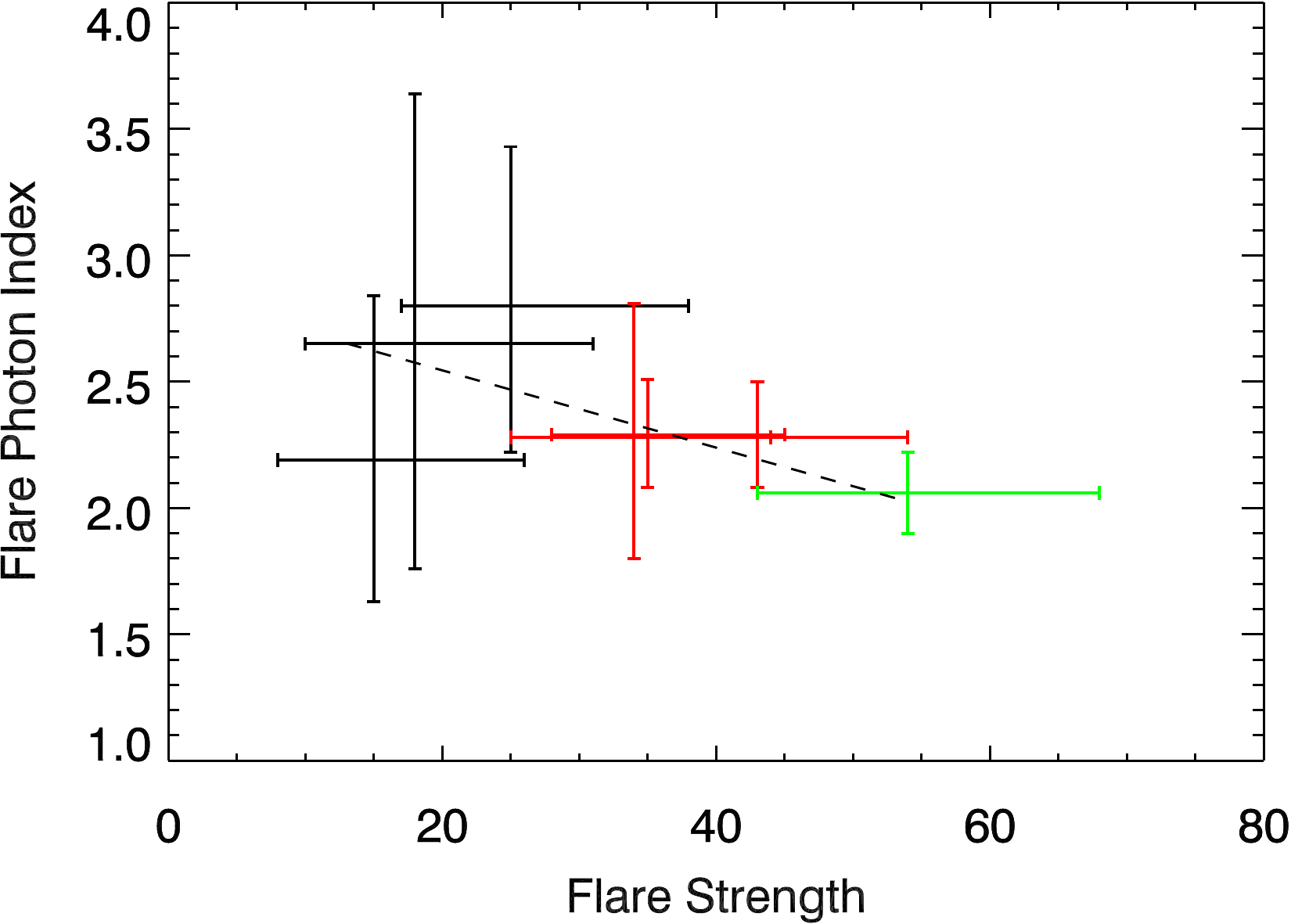}
\caption{Spectral Index vs. strength for seven \nustar\ X-ray flares with detection significance $>5\sigma$ (with 90\% error bars). The flare strength is defined as the ratio of the flare 2-10~keV unabsorbed flux and the quiescent state flux of $F_{q}(2-10\rm~keV)=0.47^{+0.04}_{-0.03}\times10^{-12}$~erg~cm$^{-2}$~s$^{-1}$. The seven flares are grouped into three sets: flares with flare strengths less than 30 times the \sgra\ quiescent flux: Nu1, Nu2 and Nu9 in black; flares with strengths higher than 30 times but lower than 50 times the \sgra\ quiescent flux: Nu3, Nu7 and Nu8 in red;  and flares with strengths higher than 50 times the \sgra\ quiescent flux: Nu6 in green. A linear fitting of the flare indices over strengths (with 90\% error bars considered) gives a slope of $a=-0.016\pm0.010$, suggesting no significant correlation between the flare spectral shape and the flare luminosity. }
\end{figure*}


\subsection{\nustar\ \sgra\ Quiescent State Emission within a 30\asec\ Radius Region}

In order to provide an upper limit to the \sgra\ quiescent state emission above 10~keV, we also measured the spectrum of the baseline emission of the 30\asec\ radius region centered on \sgra, when the supermassive black hole was in its X-ray quiescent state.
The baseline spectrum was extracted from the 2012 \sgra\ observation with \sgra\ flares removed, during which no X-ray transient activity was detected within the \nustar\ FoV.
The source is regarded as an extended source when running the \nustar\ pipeline.

The baseline X-ray emission within 30\asec\ radius of \sgra\ comes from various types of sources, including the supernova remnant Sgr A East, star clusters like IRS 13 and IRS 16, numerous X-ray point sources including G359.95$-$0.04 (a PWN candidate), and local X-ray diffuse emission \citep{Baganoff2003}. 
Due to the complexity of the baseline emission components, we used a phenomenological model to fit the spectrum.
The model we used is a combination of two thermal plasmas, a power-law and a Gaussian representing the 6.4~keV neutral Fe line, all subject to absorption {\tt Tbabs*(apec1+apec2+gaussian+power-law)}, resulting in $\chi_{nu}^{2}=1.01$ with DoF of 199 (see Figure 4).
The absorption column density is $N_{\rm H}=(1.7\pm{0.3})\times10^{23}\rm~cm^{-2}$.
The best-fit values for the temperature and abundance of the two {\tt apec} models are $kT_{1}=1.16^{+0.18}_{-0.17}\rm~keV$ with $z_{1}=2.1^{+0.7}_{-0.5}$ and $kT_{2}=7.2^{+2.1}_{-1.6}\rm~keV$ with $z_{2}=1.5^{+0.9}_{-0.6}$.
The photon index of the power-law is $\Gamma=1.64^{+0.17}_{-0.19}$.
The total absorbed flux in 2--10~keV and 10--79 keV are measured as $F_{abs,2-10}=(2.8\pm0.1)\times10^{-12}\rm~erg~cm^{-2}~s^{-1}$ and $F_{abs,10-79}=(3.7\pm0.1)\times10^{-12}\rm~erg~cm^{-2}~s^{-1}$.
The corresponding unabsorbed fluxes in these two energy bands are $F_{unabs, 2-10}=(8.2\pm0.1)\times10^{-12}\rm~erg~cm^{-2}~s^{-1}$ and $F_{unabs, 10-79}=(3.8\pm0.1)\times10^{-12}\rm~erg~cm^{-2}~s^{-1}$. 
Therefore, we derive the upper limit to the quiescent luminosity of \sgra\ above 10~keV as $L_{q, 10-79\rm~keV}=2.9\times10^{34}$~erg~s$^{-1}$.

For comparison, the unabsorbed 2--10~keV flux of \sgra\ in quiescence measured by \chandra\ is $F_{2-10}=(0.47^{+0.05}_{-0.03})\times10^{-12}\rm~erg~cm^{-2}~s^{-1}$, contributing to only 5\% of the unabsorbed 2--10 keV flux in the 30\asec\ radius region of \sgra\ we measured using \nustar. 
The thermal apec components of the spectrum mainly originates from supernova heating of the interstellar medium, coronally active stars, and non-magnetic white dwarfs (\citealp{Perez2015} and references therein).
These thermal components becomes negligible towards 20 keV, as shown in Figure 4.
The high-energy X-ray emission above 20 keV is dominated by the PWN candidate G359.95$-$0.04 \citep{Wang2006} and a newly discovered diffuse component dominating above 20 keV, which is likely an unresolved population of massive magnetic CVs with white dwarf masses $M_{WD} \sim 0.9M_{\sun}$ \citep{Revnivtsev2009, Mori2015, Perez2015, Hailey2016}.    
We compared the measured 20--40~keV \sgra\ quiescence flux with that of G359.95$-$0.04 and the hard X-ray diffuse emission.
Based on the analysis of \citet{Wang2006} on G359.95$-$0.04, its extrapolated 20--40~keV flux falling in the \nustar\ HPD circle (30\asec) is $F_{20-40, PWN}=(0.3\pm0.1)\times10^{-12}$~erg~s$^{-1}$.
According to the hard X-ray diffuse emission spatial distribution model \citep{Perez2015}, the 20--40~keV flux of this diffuse component in the inner 30\asec\ around \sgra\ is $F_{20-40, d}=(0.8\pm0.1)\times10^{-12}$~erg~s$^{-1}$.
The sum of the PWN and the hard X-ray diffuse emission 20--40~keV flux is therefore $F_{20-40, PWN+d}=(1.1\pm0.1)\times10^{-12}$~erg~s$^{-1}$, which is very close to the 20--40 flux of the inner 30\asec\ region $F_{20-40}=(1.16\pm0.05)\times10^{-12}\rm~erg~cm^{-2}~s^{-1}$ as measured using \nustar, leaving about 5\% flux from other sources.
Therefore, the high-energy flux is dominated by the contribution from the PWN candidate G359.95$-$0.04 and the hard X-ray diffuse emission.
It is reasonable to estimate that the contribution of \sgra\ is also close to 5\% above 20~keV, as it is in 2--10~keV.

\begin{figure}
\label{fig:sgraQspec}
\includegraphics[width=1.0\linewidth, angle=0]{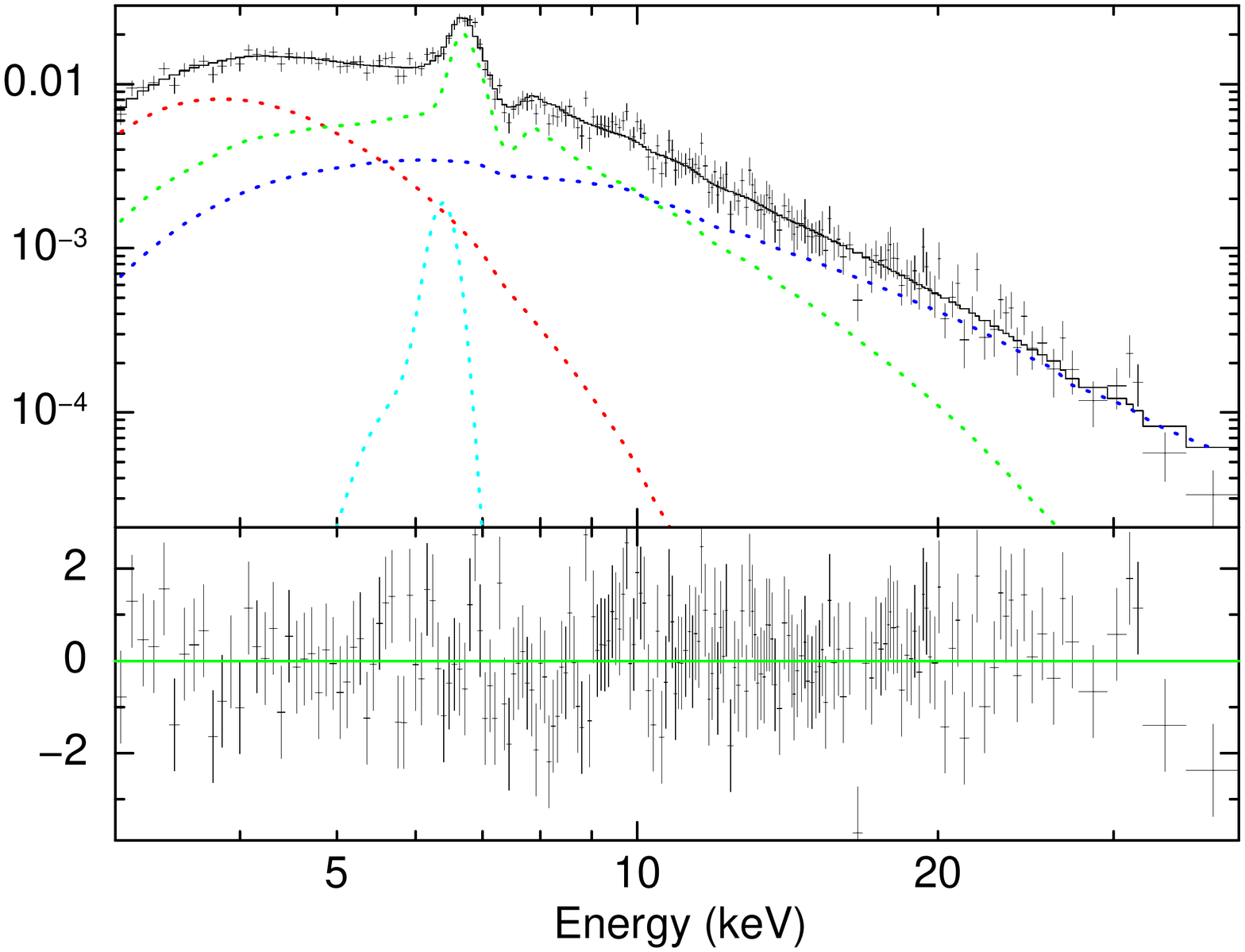}
\caption{\nustar\ FPMA spectrum for the inner 30\asec\ of the \sgra\ region during its X-ray quiescence with the best-fit model. The spectrum is well-fit with a multi-component model with individual components in different colors: two apec models with $kT_{1}\sim1.1$~keV (red) and $kT_{2}\sim6.7$~keV (green), a Gaussian for 6.4~keV neutral Fe line (cyan) and a power-law (blue), resulting in 
tbabs*(apec1+apec2+gaussian+power-law). The thermal apec components become negligible above 20~keV, where the non-thermal power-law component starts to dominate.
This spectrum can help to constrain the \sgra\ quiescent luminosity level, though it is likely dominated by a PWN candidate and diffuse X-ray emission.}
\end{figure}
   

\section{Summary and Discussion}

Using the $\sim1$~Ms \nustar\ Galactic Center observations from 2012 Fall to 2015 Spring, we searched for flaring activity from the supermassive black hole \sgra\ via Bayesian block analysis and compared our data to simultaneous X-ray observations by \chandra\ and \xmm\ to identify additional fainter events.
\nustar\ has so far captured a total of ten X-ray flares up to 79~keV.
This has allowed us to study the \sgra\ flare spectral properties with a larger flare sample in a broad X-ray energy band.

Seven flares were significantly detected at $\ge5\sigma$ confidence, with 3--79~keV luminosities ranging from $L_{3-79\rm~keV}\sim(0.7$--$4.0)\times10^{35}$~erg~s$^{-1}$, corresponding to a factor of 15--54 above the quiescent luminosity of \sgra\ (Table \ref{tab:FlareSpec}).
Four out of the seven bright X-ray flares were simultaneously detected with \chandra\ or \xmm.
Three flares are detected at lower significance due to low luminosities or limited time coverage by \nustar.

Whether there is spectral dependence on luminosity is important in discriminating and constraining both the flare radiation mechanism and understanding the physical processes behind it.
Systematic studies of \sgra\ flare data obtained by \chandra, \xmm, and \swift\ have shown no evidence for spectral/color differences among flares with different luminosities \citep{Nowak2012, Degenaar2013, Neilsen2013}.
By virtue of the broadband spectroscopy with \nustar, \citet{Barriere2014} for the first time reported a brighter flare Nu6 (O17) with a harder spectrum than a fainter flare Nu2 (J21-1).
However, with a larger \nustar\ flare dataset, we find this trend is detected below $2\sigma$, 
i.e. suggesting no significant spectral hardening for brighter flares (Figure 3).
A spectral hardening of $|\Delta \Gamma| > 1.7$ can be excluded for flares with strengths from $S=18$ to $S=54$.
As there is no strong evidence for varying spectral index from flare to flare, we accumulated all the \nustar\ flare spectra (with joint \chandra/\xmm\ spectra when available) and fit with the same model.
A simple power-law with $\Gamma=2.2\pm0.1$ provided a good fit to our current data, requiring no spectral curvature/spectral break.
The lack of variation in X-ray spectral index with luminosity and the lack of evidence for spectral curvature would point to a single radiation mechanism for the flares and is consistent with the synchrotron scenario, though the SSC model cannot be ruled out.
We note that a recent multi-wavelength study of bright flares reports a tentative detection of spectral evolution during bright flares \citep{Ponti2017}, which needs to be further tested.
Since all ten flares reported in this work are only partly captured by the \nustar\ GTIs, we are not able to verify this result using the \nustar\ dataset. 

Lastly, we show the spectrum of the inner 30\asec\ of the Galaxy when \sgra\ is in quiescence.
While the thermal components become negligible above $\sim20$~keV, a non-thermal component starts to dominate.
This is similar to the spectra from two regions at radii $r \approx 1\amin$-2\amin\ to the southwest and northeast of \sgra\ \citep{Perez2015}, 
where the dominant sources above 20~keV are likely to be an unresolved population of massive magnetic CVs.
For the inner 30\asec\ region, the dominating sources above 20~keV include not only the contribution from this massive CV population, but also a bright PWN candidate G359.95$-$0.04.
We estimate that the \sgra\ quiescence flux contributes to about 5\% of the 20--40~keV flux from the 30\asec\ region measured by \nustar.
The upper limit of the \sgra\ 10--79~keV luminosity is $L_{q, 10-79}=2.9\times10^{34}$~erg~s$^{-1}$ when the whole signal from the inner 30\asec\ is integrated.

\acknowledgements
This work was supported under NASA Contract No. NNG08FD60C, and made use of data from the \nustar\ mission, 
a project led by the California Institute of Technology, managed by the Jet Propulsion Laboratory, and funded by the 
National Aeronautics and Space Administration. We thank the \nustar\ Operations, Software and Calibration teams 
for support with the execution and analysis of these observations. This research has made use of the \nustar\ Data 
Analysis Software (NuSTARDAS) jointly developed by the ASI Science Data Center (ASDC, Italy) and the California
 Institute of Technology (USA). S.Z. ackowledges support by NASA Headquarters under the NASA Earth and Space Science 
Fellowship Program - Grant ``NNX13AM31H". J.N. acknowledges funding from NASA through the Einstein Postdoctoral 
Fellowship, grant PF2-130097, awarded by the CXC, which is operated by the SAO for NASA under contract NAS8-03060. 
G.P. acknowledges support by the Bundesministerium f{\"u}r Wirtschaft und Technologie/Deutsches Zentrum f{\"u}r Luft- und 
Raumfahrt (BMWI/DLR, FKZ 50 OR 1604) and the Max Planck Society. We thank all members of the \chandra\ \sgra\ XVP 
collaboration (http://www.sgra-star.com).

\end{document}